\documentclass{nature}

\usepackage{color}
\usepackage{amsmath} 
\usepackage{graphicx}
\usepackage{geometry}
\usepackage{textcomp}
\usepackage{amssymb}

\makeatletter
\let\saved@includegraphics\includegraphics
\AtBeginDocument{\let\includegraphics\saved@includegraphics}
\renewenvironment*{figure}{\@float{figure}}{\end@float}

\title{Ultrasensitive all-optical thermometry using  nanodiamonds with high concentration of silicon-vacancy centres and multiparametric data analysis  }

\author{Sumin Choi$^1$*, Viatcheslav N. Agafonov$^2$, Valery A. Davydov$^3$ \& Taras Plakhotnik$^1$\newline
*email: sumin.choi@uq.edu.au}

\begin{document}

\maketitle

\begin{affiliations}
 \item  School of Mathematics and Physics, The University of Queensland, QLD 4072, Australia
 \item GREMAN, UMR CNRS-7347, University F. Rabelais, 37200 Tours, France
 \item L.F. Vereshchagin Institute for High Pressure Physics,  The Russian Academy of Sciences, Troitsk, Moscow 108840, Russia
\end{affiliations}

\begin{abstract}
Nanoscale thermometry is of paramount importance to study primary processes of heat transfer in solids and is a subject of hot debate in cell biology. Here we report ultrafast temperature sensing using all-optical thermometry exploiting synthetic nanodiamonds with silicon-vacancy (SiV) centres embedded at a high concentration. Using multi-parametric analysis of photoluminescence (PL) of these centres, we have achieved an intrinsic  noise floor of about 10 mK Hz\,$^{-1/2}$, which is a thousand-fold increase in the readout speed in comparison to the current record values demonstrated with all-optical methods of comparable spatial-resolution and precision.  Our  thermometers are smaller than 250-nm across  but can detect a  0.4$^\circ$C change of temperature in a measurement taking only 0.001 second. The exceptional sensitivity and simplicity of these thermometers enable a wide range of applications such as temperature monitoring and mapping within intracellular regions and in state-of-the-art solid-state electronic nanodevices.
\end{abstract}

There is a high demand for development of ultrasmall, noninvasive, fast and reliable nanothermometry for  temperature monitoring intracellularly in vitro and in small organisms\cite{vojinovic2006real,donner2012mapping,kucsko2013nanometre}. Heat produced by living species has been known for centuries but viewed merely as a factor contributing to body temperature. Recent progress in ultra-local thermometry down to sub-cellular  spatial resolution  has challenged this concept, but the accuracy of such measurements has been questioned.  While several different types of optical nanothermometers are being developed using quantum dots\cite{yang2011quantum}, organic dyes\cite{low2008high}, nanostructures\cite{shang2013intracellular}, and luminescent proteins\cite{wong2007molecular}, a heated discussion has been sparked about the reliability of temperature measurements in cells and  the applicability of theoretical concepts based on scaling laws and macroscopic thermodynamics to the complex intracellular interior\cite{baffou2014critique,celular_therm_pro,celular_therm_pro2, baffou2015reply}.  Reduced size of solid state devices with faster switching speeds has also increased the need for accurate and fast temperature measurement  with sub-micron resolution especially during real-time operation\cite{heat_nanodevices}. Detection of local temperature is extremely important due to the complexity of the heat transport on the nanoscale where the conventional Fourier's law breaks down\cite{phonons_difusion}. Thus, development of fast nanothermometers free of artefacts is highly crucial. 

During past decades, a very active field  of non-invasive optical nanothermometry has emerged\cite{jaque2012luminescence,brites2012thermometry} where nanodiamonds imbedded with  colour centres have attracted much interest\cite{neumann2013high,sekiguchi2018fluorescent,kucsko2013nanometre} due to their remarkable optical properties\cite{aharonovich2011diamond}, biocompatibility\cite{zhu2012biocompatibility}, and uncomplicated surface functionalization\cite{liu2004functionalization}. Although a large number of colour centres exists in diamond, nitrogen vacancy (NV)  combined with optically detected magnetic resonance (ODMR) technique have dominated research\cite{plakhotnik2015all,alkahtani2017nanometer,alkahtani2018fluorescent}.  The limits in precision are conventionally expressed in terms of a noise floor and have been reported as small as 130 mK Hz\,$^{-1/2}$ and 25 mK Hz\,$^{-1/2}$ using single NV centres\cite{neumann2013high}  and magnetic nanoparticle hybrid nanodiamond\cite{wang2018magnetic}, respectively. However, ODMR measurements require microwave radiation which can induce sensitive temperature changes at cellular levels\cite{de2005microwaves} and are quite complicated in implementation. Both factors are undesirable for biological applications. 

Alternatively, a simple all-optical method of temperature measurements with NV-centres has been demonstrated recently\cite{plakhotnik2015all} using a feature in their luminescence spectra called a zero phonon line (ZPL). However the strength of the ZPL in NV-centres  is very small and this limits the noise floor to 0.3 K\,Hz$^{-1/2}$.  Other centres with stronger ZPL transition such as silicon vacancy (SiV)\cite{nguyen2018all}, germanium vacancy (GeV)\cite{fan2018germanium}, and tin vacancy (SnV)\cite{alkahtani2018tin} have emerged as promising candidates for temperature sensing but their reported precisions are similar to the value obtained with NV-centres: 0.5\,K\,Hz\,$^{-1/2}$ (SiV in 200-nm crystal) and 0.3\,K\,Hz\,$^{-1/2}$ (GeV and SnV in a bulk crystal). 

 Here we show ultrasensitive all-optical temperature measurements using photoluminescence (PL)  of SiV centres in  diamonds smaller than 250 nm and  demonstrate intrinsic noise floor of about 10\,mK Hz\,$^{-1/2}$, a dramatic improvement of the current record values. This reduction of the noise is equivalent to a thousand-fold decrease of the acquisition time for a given precision of thermometry. The factor of 1000 has two contributions. A part on the order of 100 is due to the improvement in the crystal growth process which allows for a much higher concentration of SiV;  the remaining order of magnitude is obtained by developing a novel method of data analysis. 

\begin{figure}[htbp]
   \centering
   \includegraphics [width=10cm]{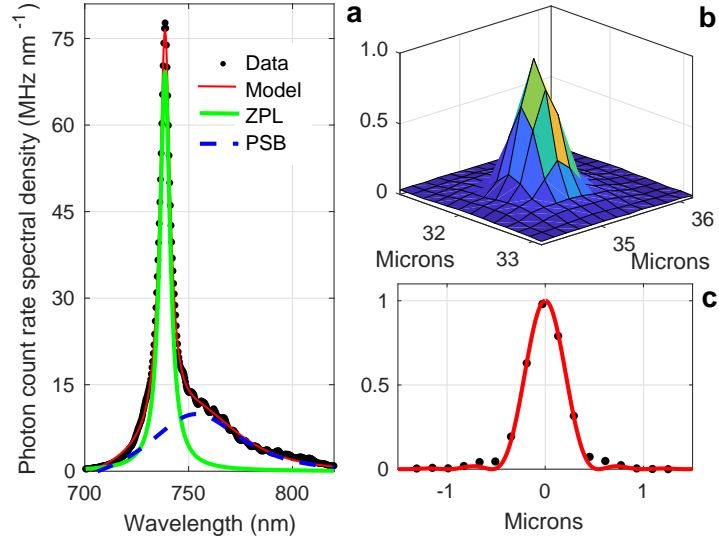} 
   \caption{ (\textbf{a}) PL spectrum of a diamond crystal with SiV centres. Black dots show experimental data. At 50-mW laser power (0.55 MW\,cm$^{-2}$ power density at the location of the crystal) the detected photon count rate is about 0.8 GHz. The red curve is the best fit of the function defined by Eq.(\ref{eq:spectrum}) to the experimental data. The green and blue curves show the ZPL and  the PSB respectively.  Panel (\textbf{b}) shows an optical image of a typical bright crystal with vertical scale proportional to the luminescence intensity. Panel (\textbf{c}) shows a cross-section of the image in (\textbf{b}) scaled vertically and shifted horizontally to overlap with a theoretical point spread function (red line) of the microscope objective.  }
   \label{fig:PL}
\end{figure}

Details of optical measurements are presented in \textbf{Methods}. An example of a spectrum is shown in Fig.\,\ref{fig:PL}a.   Luminescence spectra of SiV centre can be split into two distinctive bands, a strong ZPL at around 740 nm and the phonon side band (PSB), an approximately 25-nm broad feature to the right of the ZPL.  We did not observe significant spectral signatures of other defect centres in the measured nanodiamonds. This is important  because presence of different colour centres with overlapping spectra\cite{choi2018varying} results in complexity of the data analysis.  In the present case, SiV spectra  are accurately modelled with the function  $\Phi(\lambda)$ below, a sum of two Lorentzians and background $B(\lambda)$ (assumed to be a straight line) multiplied by  an overall scaling factor $R_0$.  
\begin{equation}\label{eq:spectrum}
\Phi(\lambda)=R_0\left[A_\mathrm{zpl}\frac{ \gamma^2}{4(\lambda-\lambda_\mathrm{zpl})^2+\gamma^2}+\frac{\Gamma^2}{4(\lambda-\lambda_\mathrm{psb})^2+\Gamma^2}+B(\lambda)\right]
\end{equation} 
 Generally, the temperature of the crystal affects the widths of the ZPL and of the PSB  ($\gamma$ and $\Gamma$, respectively), the peak positions of each Lorentzian ($\lambda_\mathrm{zpl}$ and $\lambda_\mathrm{psb}$), and the relative amplitude of the ZPL ($A_\mathrm{zpl}$). Some of the parameters appear to be more sensitive than others. These dependencies are used for temperature sensing.  
 
 Fig.\,\ref{fig:PL}b shows a typical optical image of a crystal,  direct evidence that the data are collected from a small object.  A vertical cross-section of the image in Fig.\,\ref{fig:PL}b is shown in Fig.\,\ref{fig:PL}c together with a diffraction limited point spread function (PSF) for a microscope objective with $\mathrm{NA}=0.9$  and at the wavelength of 740 nm (all as in the experiment). The agreement between the experimental data and the theoretical PSF indicates that the size of the light emitting object is not larger than 250 nm. 

The most probable values of  $\gamma$, $\Gamma$, $\lambda_\mathrm{zpl}$, $\lambda_\mathrm{psb}$, and $A_\mathrm{zpl}$  have been found from experimental spectra using conventional least-squares fitting. It is important to note that although this method has theoretical justifications (a normal distribution of the errors in the intensity at every spectral point and statistical independences of these errors), such a routinely used "fitting" may result in unphysical correlations between the fitted parameters. For example, fitting a single Lorentzian to a spectrum creates correlation between the peak value of the Lorentzian and its width\cite{fit_corr} (see the reference for other examples). Such correlations are not caused by mutual physical dependence of the fitted parameters but results from  the dependence of the minimal value of $\chi^2$-function (the criteria for the best fitting) on these parameters.  These correlations are important and will be discussed in relation to temperature measurements. 

\begin{figure}[htbp] 
   \centering
   \includegraphics  [width = 7cm] {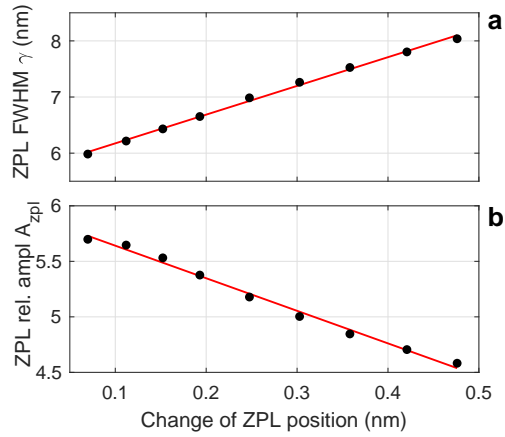}
   \caption{Values of (\textbf{a}) $\gamma$ and (\textbf{b})  $A_\mathrm{zpl}$ versus the change of $\lambda_\mathrm{zpl}$. The gradients of the red lines are are $5.0\pm0.2$ and $3.0\pm0.2\,\text{nm}^{-1}$  respectively. The change of $\lambda_\mathrm{zpl}$ is shown relative to the value obtained by extrapolating $\lambda_\mathrm{zpl}$  to zero laser power (in this case the temperature of the crystals equals 21$^\circ$C, the lab temperature). In the shown range of data, the temperature varies from 25$^\circ$C to 54$^\circ$C.}
   \label{fig:par_depend}
\end{figure}

A series of SiV-spectra has been measured at different excitation laser powers. The resulting values of  $\gamma$ and $A_\mathrm{zpl}$   vs $\lambda_\mathrm{zpl}$ are plotted in Fig.\,\ref{fig:par_depend}. It can be seen that the broadening of the ZPL is about 5 times larger than the corresponding shift. A similar trend has been reported for dependence of the ZPL on temperature\cite{dragounova2017determination}. The heating is related to a high concentration of silicon in the primary mixture for crystal growth (see \textbf{Methods}). The excess of silicon forms approximately spherical objects which are mixed with nanodiamonds in the final product (see Supplementary Information). 
Silicon spheres (some of them partially oxidized) can be dissolved and removed from the product but here they were used for heating. The absorption of light by these impurities increases the temperature of  nearby diamonds and thus their thermal properties can be investigated without need of an external heating element. The temperature  could be simply varied by changing the power density of the exciting laser light at the location of the spheres. The laser-induced heating is most significant when the microscope objective operates in a wide field excitation mode. In a confocal mode when the laser light is tightly focussed, this heating is reduced by an order of magnitude for the same power density at the location of the diamond crystal. This proves that the heating is not caused by absorption of light in the diamond crystal itself. 

The temperature change of a specific diamond crystal can be accessed using one of the parameters defining the luminescence spectrum of SiV. We have used the peak wavelength of the ZPL  which has been studied several times and for which a consistent value of its temperature sensitivity $s_{\lambda_\mathrm{zpl}}\equiv \mathrm{d}\lambda_\mathrm{zpl}/\mathrm{d}T\approx 0.012\,\text{nm\,K}^{-1}$   has been reported\cite{nguyen2018all, lagomarsino2015robust}.  Temperature sensitivity of all other parameters could be obtained using the sensitivity of the $\lambda_\mathrm{zpl}$ as a reference. For example, sensitivity of the ZPL width and its relative amplitude can be estimated as about $s_\gamma=0.062 \text{nm\,K}^{-1}$ and $s_\mathrm{A}=0.037\,\text{K}^{-1}$  respectively using the data shown in Fig.\,\ref{fig:par_depend}.  

Due to the correlation mentioned above, even temperature-independent parameters may unfavourably affect the precision of  the temperature estimate by transferring their noise to temperature-dependent parameters. The multiparametric analysis eliminates the effect of correlation to achieve the best outcome from the measurements.  A general approach is described in \textbf{Methods}.  In this case we have used  the following linear combination of three parameters as a single temperature sensing factor 
\begin{equation}\label{eq:G3}
G_3=\gamma+w_\lambda \lambda_\mathrm{zpl}- w_\mathrm{A}A_\mathrm{zpl}
\end{equation}
Addition of other parameters to the linear combination has insignificantly improved the results. The procedure of finding the optimal values of the weighting factors is outlined below.  

\begin{figure}[htbp]
   \centering
   \includegraphics  [width = 12cm] {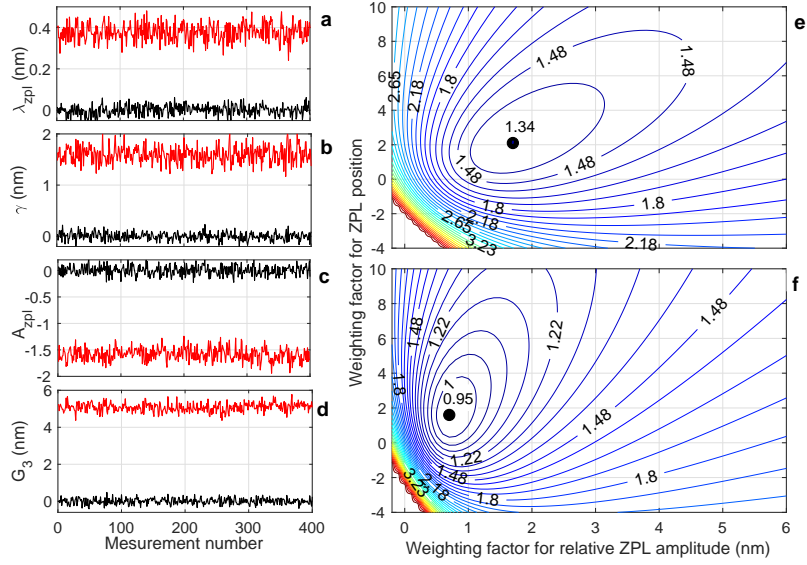}  
   \caption{Multiparametric data analysis at the measurement time of 0.1 ms. On the left: Temperature sensing using a single parameter.  The black lines are recorded at  laser power  of 6\,mW (estimated temperature  298 K) and the red lines are recorded at 50 mW ($T\approx327\,\text{K}$). All curves are shifted vertically so that the average values at the lower temperature are zero. (\textbf{a}) ZPL peak position $\lambda_\mathrm{zpl}$, (\textbf{b}) ZPL width $\gamma$, (\textbf{c}) ZPL relative amplitude  $A_\mathrm{zpl}$, (\textbf{d}) $G_3$ with weighting factors optimized for the best precision.  On the right: The contour plots show $\sigma_\mathrm{T}$ in units of K as a function of $w_\lambda$ and $w_\mathrm{A}$. Panel (\textbf{e}) is the standard deviation calculated using 50-mW data and (\textbf{f}) obtained using spectra simulated with experimental parameters  substituted into Eq.\,(\ref{eq:spectrum}) and added Poisson noise. The black dots locate the weighting factors for the best precision of the temperature. }
   \label{fig:G_opt}
\end{figure}

Panels \textbf{a-c} in Fig.\,\ref{fig:G_opt} show $\lambda_\mathrm{zpl}$, $\gamma$ and $A_\mathrm{zpl}$   determined for two sets of  400 spectra each recorded with 0.1-ms integration time at two temperatures. The best precision was determined at the higher temperature (also higher laser power) while data at the lower temperature have been used for thermal sensitivity calibration. The optimal values of $w_\lambda \approx 2.1$ and $w_\mathrm{A} \approx 1.7\,\text{nm}$  (note that factor $w_\mathrm{A}$ has units while $w_\lambda$ is dimesionless) were determined empirically, as shown in Fig.\,\ref{fig:G_opt}\,(\textbf{e}, \textbf{f}) by plotting the values of $\sigma_\mathrm{T}$ versus $w_\lambda$ and $w_\mathrm{A}$.   The three-parametric sensing method improves the precision as can be seen in  Fig.\,\ref{fig:G_opt}\textbf{d}.  Given the measuring (integration) time $t_\mathrm{m}=0.1\,\text{ms}$, the minimal standard deviation of $1.34\pm0.1$\,K corresponds to the noise floor $\eta_\mathrm{T} \equiv \sigma_\mathrm{T}t_\mathrm{m}^{1/2} = 13.4\pm1\,\text{mK\,Hz}^{-1/2}$.  A single-parameter (ZPL width) case corresponds to the point with co-ordinates $w_\lambda =0$ and $w_\mathrm{A} =0$ in Fig.\,\ref{fig:G_opt}\textbf{e}. It can be seen that the multiparametric analysis improves  the precision by a factor of approximately 2.5 in this case  in comparison to the results achieved using ZPL width or ZPL peak position alone. Equivalently, it  decreases of the measurement time  by 6-fold if the precision is kept at a fixed value. Monte Carlo simulations of the spectra with Poissonian noise result in the minimal standard deviation of 0.95\,K (see Fig.\,\ref{fig:G_opt}f). This value is approximately $\sqrt{2}$ times smaller than the 1.34\,K due to the noise excess factor\cite{EMCCD} of electron-multiplying  CCDs. If the data were taken without electron multiplication (for example, with a sCMOS camera)  the noise floor would be reduced to about 10\,mK\,Hz\,$^{-1/2}$. The  precision is inversely proportional to the square-root of the integration time (see below) and, for example,  improves to $(0.4\pm0.1)$\,K  at 1 ms. Previously, a similar precession with diamond colour centres has been obtained\cite{alkahtani2018tin} only after integration for 1\,s. 

\begin{figure}[htbp]
   \centering
   \includegraphics [width=10cm]{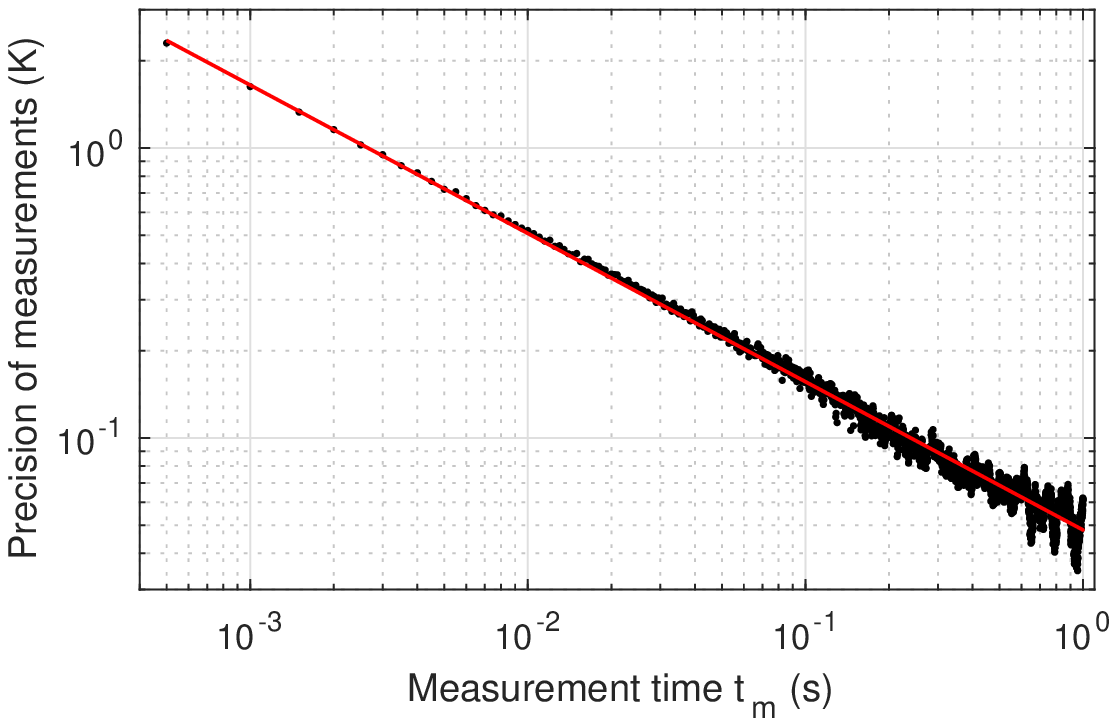} 
   \caption{The precision $\sigma_\mathrm{T}$ of temperature measurement at 6-mW laser power (corresponds to 4\,K above the lab temperature) at different measurement times $t_\mathrm{m}$. The laser power density at the location of the nanodiamond and the detected photon rate are approximately 70 kW\,cm$^{-2}$  and 25\,MHz. $G_2 =\gamma- 0.42\,\text{nm} \times A_\mathrm{zpl}$ is used for thermometry. The straight line corresponds to the theoretical dependence $\sigma_\mathrm{T} \propto 1/\sqrt{t_\mathrm{m}}$, where the proportionality constant is the noise floor $\eta_\mathrm{T}= 49\pm1\,\text{mK\,Hz}^{-1/2}$. It is confirmed by numerical simulations (not shown) that due to the excess noise of the EMCCD this value is $\sqrt{2}$ times larger than it should be theoretically for the given count rate and Poisson statistics of the noise. }
   \label{fig:mesuring_time}
\end{figure}

 As an ultimate demonstration of our method we have recorded 20\,000 spectra with 0.5-ms integration time and power density of the exciting laser reduced to about 70 kW\,cm$^{-2}$, a value more common for intracellular studies. This set of  data has been used to obtain precision of the temperature measurements at different integration times. For example, one can obtain 100 spectra with effective integration times of 0.1\,s and calculate the corresponding  standard deviation by splitting the entire data set into 100 subsets and averaging each. Results obtained in this way  for integration times up to 1\,s, where the precision is better than 50 mK, are shown in Fig.\,\ref{fig:mesuring_time}.    
Long integration times  undesirably affect $\lambda_\mathrm{zpl}$ more than other parameters. The 50-mK change of temperature shifts the ZPL by $6\times10^{-4}\,\text{nm}$ spectrally and  by about 30\,nm on the CCD sensor. Due to the magnification of the microscope  objective, a 1-nm shift of the nanocrystal position displaces ZPL comparably with the thermal shift. For this reason, $\lambda_\mathrm{zpl}$ has been excluded from Eq.\,(\ref{eq:G3}).  

It is interesting to compare the achieved precision of nanothermometers with the fundamental limit set by thermodynamics on the concept of temperature. Due to the small size, the temperature of a nanodiamond fluctuates and these fluctuations can be characterized by a standard deviation $\sigma_\mathrm{th}$ expressed as
\begin{equation}
\sigma_\mathrm{th}=T\sqrt{\frac{k_\mathrm{B}}{vc_\mathrm{v}}}
\end{equation}
where $k_\mathrm{B}$ is the Boltzman's constant, $v$ is the volume of the crystal, and $c_\mathrm{v}$ is the constant volume heat capacity of diamond per unit volume\cite{Landau_stat}.  $\sigma_\mathrm{th}$ calculated for a 250-nm sphere of diamond at $T\approx 300\,\text{K}$ is about 10\,mK  and is close to the precision of our thermometers. The characteristic timescale of these fluctuations ($\tau_\mathrm{rel}$) is defined by the heat exchange between the crystal and its environment. A nanodiamond immersed in a homogeneous media has $\tau_\mathrm{rel}$ on the order of $ a^2/\alpha$, where $a$ is the radius of the crystal and $\alpha$ is the thermal diffusivity of the media. The observed thermal fluctuations equal $\sigma_\mathrm{th} (\tau_\mathrm{rel}/t_\mathrm{m})^{1/2}$ and therefore the  limit on the noise floor set by thermal fluctuations is $\eta_\mathrm{th,T} = \sigma_\mathrm{th} (\tau_\mathrm{rel})^{1/2}\approx T[k_\mathrm{B}/(a c_\mathrm{v}\alpha)]^{1/2}\approx 6\text{\,\textmu K\,Hz}^{-1/2}$ for a 250-nm diamond in water. 

In conclusion, we present an intrinsic 10\,mK Hz$^{-1/2}$ noise floor in all-optical thermometry using  diamonds smaller than 250-nm doped with SiV centres.  A high SiV concentration and multiparametric analysis enable  about 1000 times faster measurement if compared to any reported up to date an all-optical detection scheme with sub 250-nm spatial resolution. The value for the noise floor is valid for integration times as long as 1\,s.

\begin{methods}
\subsection{Sample preparation}
Nanodiamonds with SiV impurity-vacancy colour centres were obtained by high pressure - high temperature (HPHT) treatment of the catalyst metals-free hydrocarbon growth system based on homogeneous mixtures of naphthalene - $C_{10}H_{8}$  (Chemapol) and tetrakis(trimethylsilyl)sylane – $C_{12}H_{36}Si_{5}$ (Stream Chemicals Co.), which was used as the doping component\cite{davydov2014production}. Cold-pressed tablets of the initial mixture (5 mm diameter and 4 mm height) were placed into a graphite container, which simultaneously served as a heater of the high-pressure toroid-type apparatus. The experimental procedure consisted of loading the high-pressure apparatus to 8.0 GPa at room temperature, heating the sample to the temperature of diamond formation (1400°C), and short (10 s) isothermal exposure of the sample at this temperature. The obtained high-pressure states have been isolated by quenching to room temperature under pressure and then complete pressure release. The recovered diamond materials have been characterized under ambient conditions by using X-ray diffraction, Raman spectroscopy, scanning and transmission electron microscopies (SEM and TEM). The nanodiamonds are dispersed with de-ionised water and treated with 120\,W power ultrasound for 30 seconds. Then the nanodiamonds are drop-cast on a cover glass for optical characterisation. The mean size of nanodiamonds is about 70 nm, but much larger crystals are present in the sample.

\subsection{Optical measurements}
The optical measurements are made with a home-built confocal microscope at room temperature. Off-resonant excitation light from a continuous-wave 532-nm laser is focused onto the sample through an air objective $\times100$ and NA = 0.90 (Nikon). The emission passes through a dielectric filter to reject the excitation light, then is collected into an EMCCD (iXon Andor). The apparatus can work in imaging or spectroscopic modes (with Acton 2300i grating spectrometer) and use wide-field or confocal illumination. In the confocal mode, the laser light intensity in the focal spot is proportional to $\exp(-r^2/r_0^2)$, where $r$ is the distance from the centre of the spot (and the location of the investigated crystals) and $r_0 = 1.2$\,\textmu m. This can be used to estimate the power density of light at the centre of the spot to be 11\,MW\,cm$^{-2}$ at 1-W output of the exciting laser (taking into account about 50\% loss of power on all optical elements of the experimental setup). 

\subsection{Data analysis} For simplicity we consider a case of two parameters $X=\overline{X}+x$ and $Y=\overline{Y}+y$ deduced from analysis of experimental data, where $x$ and $y$ are random errors with zero means and standard deviations $\sigma_x$ and $\sigma_y$ respectively. We assume that $\overline{X}$ and $\overline{Y}$ depend linearly  on a physical quantity of interest, for example, the temperature. That is $\overline{T}=s_X\overline{X}$ and $\overline{T}=s_Y\overline{Y}$. If $x$ and $y$ are not correlated, the best estimate for $T$ can be obtained using $Q$, a linear combination of $X$ and $Y$ defined as follows 
\begin{equation}\label{eq:define_Q}
Q\equiv\frac{X}{s_X\sigma^2_x}+\frac{Y}{s_Y\sigma^2_y}=\left(\frac{1}{s^2_X\sigma^2_x}+\frac{1}{s^2_Y\sigma^2_y}\right)T
\end{equation}
For example, the estimation for $T$ easily derived from Eq.\,(\ref{eq:define_Q}) 
\begin{equation}
T=\frac{Q}{\frac{1}{s^2_X\sigma^2_x}+\frac{1}{s^2_Y\sigma^2_y}}=\frac{\frac{s_XX}{s^2_X\sigma^2_x}+\frac{s_YY}{s^2_Y\sigma^2_y}}{\frac{1}{s^2_X\sigma^2_x}+\frac{1}{s^2_Y\sigma^2_y}}
\end{equation}
 is an average of two measurements $T_X\equiv s_XX$ and $T_Y\equiv s_YY$, weighted by the inverse squares of the corresponding standard deviations, in agreement with elementary textbooks on data analysis. 

Correlation between $x$ and $y$ (physical or resulting from a fitting procedure)  are quantified by the expression $\overline{xy}=\rho\sigma_x\sigma_y$, where $-1\le\rho\le1$.  Uncorrelated variables $X'$ and $Y'$ can be obtained by rotation in the $(X,Y)$-plane. 
\begin{align}\label{eq:rotation}
X' \equiv X \cos\phi-Y \sin\phi\\
Y'\equiv X \sin\phi+Y \cos\phi \notag
\end{align}
The angle is determined by the condition $\overline{x'y'} =0$. With help of Eq.\,(\ref{eq:rotation}) one gets
\begin{equation}
\overline{x'y'}=(\sigma^2_x-\sigma^2_y)\cos\phi\sin\phi+(\cos^2\phi-\sin^2\phi)\overline{xy} 
\end{equation} 
so that $\overline{x'y'} =0$ if $\tan2\phi=2\overline{xy}/(\sigma^2_y-\sigma^2_x)$.  
One can solve Eq.\,(\ref{eq:rotation}) for $X$ and $Y$ and then substitute $X= s_XT_X$ and $Y=T_Ys_Y$ to find that
\begin{align}\label{eq:rot_error}
T_{X'}=\frac{X'}{s_X\cos\phi-s_Y \sin\phi}\\
T_{Y'}=\frac{Y'}{s_X \sin\phi+s_Y \cos\phi}\notag
\end{align}
where the prime in the subscripts of $T$ indicates that these estimates are obtained using $X'$ and $Y'$. Because the errors $x'$ and $y'$ are uncorrelated, one can use Eq.\,(\ref{eq:define_Q}) to find the best estimate for $T$. The standard deviation of this estimate reads
\begin{equation}\label{eq:sigma_T}
\sigma'_T=\left(\frac{(s_X \cos\phi-s_Y \sin\phi)^2}{\sigma^2_{x'}}+\frac{(s_X \sin\phi+s_Y \cos\phi)^2}{\sigma^2_{y'}}\right)^{-1/2}
\end{equation}

Without a loss of generality we take $\sigma^2_x=\sigma^2_y\equiv \sigma^2$ (the equality can be achieved by scaling) and find that $\overline{x'y'} =0$ if $\cos^2\phi=\sin^2\phi$. The variances of $x'$ and $y'$ then read 
\begin{equation}
\sigma^2_{x'}=\sigma^2(1\mp\rho) \text{ and }\sigma^2_{y'}=\sigma^2(1\pm\rho)
\end{equation}

The value of $\sigma'_T$ is smaller than the marginal value $\sigma_T= s^{-1}_X\sigma_x$ even if $Y$ is temperature-independent, that is  if $s_Y=0$. In such a case Eq.\,(\ref{eq:sigma_T}) reads $\sigma'_T=  s^{-1}_X\sigma(1-\rho^2)^{1/2}$.   If $\rho=0$, correlation between $x$ and $y$ is absent, and the equality $\sigma'_T=\sigma_T$ holds. 

Because  Eq.\,(\ref{eq:define_Q}) and Eq.\,(\ref{eq:rotation}) are linear, the best estimate for $T$ can be obtained using a single parameter $G_2$ which is a linear combination of $X$ and $Y$, that is $G_2=X+wY$. Because the overall scaling of $G_2$ is of no significance, the factor in front of $X$ can be set to 1. This idea is easily generalized to several sensing parameters. For example,  
\begin{equation}
G_n=X+\sum_{k=1}^n w_kY_k
\end{equation}
gives  the most precise value of temperature sensing with $n$ parameters.  All the factors $w_k$ can be determined theoretically (using Eq.\,(\ref{eq:define_Q}) and Eq.\,(\ref{eq:rotation}) if sensitivities and cross-correlations are known) or experimentally by optimizing $w_k$  to minimize the standard deviation of the measured quantity.

\end{methods}



\begin{addendum}
 \item The work was supported by Human Frontier Science Program, Grant RGP0047/2018.  V.\,D.  thanks the Russian  Foundation for Basic Research (Grant 18-03-00936) for financial support. 
 \item[Author contributions]  T.\,P.  managed the project. S.\,C. and T.\,P. conducted the optical experiments and analysed data. V.\,A. and V.\,D. fabricated the nanodiamonds and conducted structural analysis. All authors contributed to the scientific discussion and reviewed the manuscript. 
 \item[Competing Interests] The authors declare that they have no
competing financial interests.
 \item[Additional information] Supplementary information is available for this paper.
 \item[Correspondence] Correspondence and requests for materials should be addressed to Associate Professor Taras Plakhotnik (email:taras@physics.uq.edu.au).
\end{addendum}

\end{document}